\documentclass[useAMS,usenatbib]{mn2e}
\usepackage{graphicx}
\usepackage{amssymb}
\title[Average GRB X-ray flaring activity]{On the average Gamma-Ray Burst X-ray flaring activity}
\author[R. Margutti et al.]{R. Margutti$^{1,2}$\thanks{E-mail: raffaella.margutti@brera.inaf.it (RM)},
G. Bernardini$^{1}$, R. Barniol Duran$^{3,4}$,  C. Guidorzi$^{5}$, R.~F. Shen$^{4}$, \and G. Chincarini$^{1,2}$  \\
$^{1}$INAF Osservatorio Astronomico di Brera, via Bianchi 46, Merate 23807, Italy \\
$^{2}$ Univerisit\'a Milano Bicocca, Dip. Fisica G. Occhialini, P.zza della Scienza 3, Milano 20126, Italy\\
$^{3}$Department of Physics, University of Texas at Austin, Austin, TX 78712, USA\\
$^{4}$Department of Astronomy, University of Texas at Austin, Austin, TX 78712, USA\\
$^{5}$Dipartimento di Fisica, Universit\`a di Ferrara, via Saragat 1, 44122 Ferrara, Italy\\}

\begin{document}

\date{Accepted 2010 August 10. Received 2010 August 6; in original form 2010 July 14}
\pagerange{\pageref{firstpage}--\pageref{lastpage}} \pubyear{2010}
\maketitle

\label{firstpage}

\begin{abstract}
Gamma-ray burst X-ray flares are believed to mark the late time activity of the central engine. 
We compute the temporal evolution of the average flare luminosity $\langle L \rangle$ in the common 
rest frame energy band of 44 GRBs taken from the large \emph{Swift} 5-years data base. 
Our work highlights the importance of a proper consideration of the threshold of detection of 
flares against the contemporaneous continuous X-ray emission.
In the time interval $30\,\rm{s}<t<1000\,\rm{s}$ we find $\langle L \rangle\propto t^{-2.7\pm 0.1}$;
this implies that the flare isotropic energy scaling is $E_{\rm{iso,flare}}\propto t^{-1.7}$. The decay of the 
continuum underlying the flare emission closely tracks the average flare luminosity evolution,
with a typical flare to steep-decay luminosity ratio which is $L_{\rm{flare}}/L_{\rm{steep}}=4.7$:
this suggests that flares and continuum emission are deeply related to one another.
We infer on the progenitor properties considering different models. 
According to the hyper-accreting black hole scenario, the average flare luminosity scaling can be obtained 
in the case of rapid accretion ($t_{\rm{acc}}\ll t$)  or when the last $\sim 0.5 M_{\sun}$ of 
the original $14 M_{\sun}$ progenitor star are accreted. Alternatively, the steep $\propto t^{-2.7}$ 
behaviour could be triggered by a rapid outward expansion of an accretion shock in the material feeding a 
convective disk. If instead we assume the engine to be a rapidly spinning magnetar, then  its rotational
energy can be extracted to power a jet whose luminosity is likely to be between 
the monopole ($L\propto e^{-2t}$) and dipole ($L\propto t^{-2}$) cases.
In both scenarios we suggest the variability, which is the main signature of the flaring activity, to be established
as a consequence of different kinds of instabilities.

\end{abstract}

\section{Introduction}
Observations of  long gamma-ray bursts (GRBs) by a number of spacecrafts during the last decades  revealed 
that  their $\gamma$-ray prompt emission is over after $\sim10-100$ s. However, after the launch of  \emph{Swift} (\citealt{Gehrels04})
evidence is accumulating that the central engine of the GRB sources could still be active hours after the main burst.
This is mainly linked to the discovery of large amplitude, episodic re-brightenings with typical $\Delta t/t \sim 0.2$ in the 
$\sim 33\%$ of X-ray afterglows: the X-ray flares (see \citealt{Chincarini10}, C10 hereafter, for a recent  compilation). 
The temporal properties of X-ray flares  make it difficult to interpret the observed emission in the framework of the 
external shock scenario (see e.g. \citealt{Lazzati07}). Moreover, the strict analogy found by \cite{Margutti10b} (M10 hereafter) 
between the temporal and spectral behaviour of X-ray flares and prompt emission pulses strongly suggests a common, internal 
origin. The direct implication is that flares directly trace the activity of the central engine.  

To account for the details of the flare emission, GRB central engines (CE) are required to have the following basic properties
(C10, M10 and references therein):
\begin{enumerate}
	\item \emph{Energetics}. The average flare fluence is $\sim10\%$ the prompt 15-150 keV fluence.  However, CEs
	should be able to produce X-ray flares  whose  fluence is able to compete in some cases (e.g. GRB\,050502B) with the prompt 
	$\gamma$-ray fluence. The typical early-time flare isotropic energy is $\sim10^{51}\,\rm{erg}$. The flare peak luminosity decays
	as $L_{\rm{peak}}\propto t^{-2.7}$ for $t<1000$ s.
	\item \emph{Lifetime}. CEs are required to be long-lived and in particular to be active up to $t\sim 10^4-10^5$ s after the end of 
	the prompt emission. The probability of a CE to re-start decays with time.
	\item \emph{Variability}.  CEs are required to store and release energy in the form of erratic, short-lived episodes of emission giving rise 
	to flares whose duration linearly grows with time. 
\end{enumerate}

Ideas on how to revive the CE have been explored by a number of authors:
flares could be the result of a two-stage stellar collapse leading to core fragmentation and subsequent accretion (\citealt{King05}).
Alternatively, the fragmentation of the outer part of an hyper-accreting disk into rings of material as a result of gravitational instabilities
could potentially lead to large-amplitude variations of the CE output (\citealt{Perna06}). 
However, \cite{Piro07} noticed that tidal disruption of a fragment and accretion by the central black hole might be too rapid to account 
for the durations of observed flares.
 \cite{Proga06} proposed that magnetic flux accumulated around the accretor could stop and then restart the CE 
release of energy possibly giving rise to flares. \cite{Dai06} invoked the presence of a differentially rotating millisecond pulsar
where magnetic reconnection events are responsible for the observed flare emission. Magnetic reconnection is also suggested
by  \cite{Giannios06} in the framework of the internal plus external shock scenario with strongly magnetised ejecta: in this case, 
no late-time CE activity is needed  and flares come from the revival of MHD instabilities triggered 
by the deceleration of the magnetised ejecta in the external medium.

The entire list of flare models basically refers to two different CEs: an accretion disk around a newly formed black hole
(e.g. \citealt{MacFadyen99}) or a rapidly rotating magnetar (e.g. \citealt{Usov92}; \citealt{Thompson94}; \citealt{Wheeler00};
\citealt{Thompson04}). 
In magnetar models the ultimate source
of energy is represented by the magnetar spin and flares are likely to be connected to episodes of magnetic energy dissipation.
In the case of accretion models, the source of energy powering the flare emission is possibly of gravitational origin. However,
magnetic effects could still play the key role, deeply affecting the accretion dynamics and release of energy (e.g. 
\citealt{Fan05}; \citealt{Proga06}; \citealt{Lei09}; \citealt{Zhang09}).  X-ray flares mark the late time activity of the CE
after the main power emission and provide information about the way the CE power is progressively declining.
In the following we show that this information turns out to be of fundamental importance in constraining 
existing CE models (accretion or magnetar).


The main goal of this work is to compute the evolution of the average luminosity of the flaring component with time, taking
advantage from the large \emph{Swift} 5-years data-base. A variety of physical mechanisms which might properly account for the new 
episodes of energy release are discussed in the context of both accretion and magnetar CE models.
A previous attempt in this direction was made by \cite{Lazzati08} (L08 hereafter):  from a sample of 9 long GRBs
and 1 short GRB (which was the largest sample with known redshift available at that time) these authors showed that the average
flare luminosity follows a power-law decay $\langle L \rangle_{\rm{L08}}\propto t^{-1.5\pm 0.2}$ and concluded that accretion onto a
compact object could reasonably account for the $\langle L \rangle$ decay. In the present work we aim at presenting a substantial 
update of the L08 sample, including 44 GRBs with redshift; the flare luminosity of each GRB is computed in the rest frame energy band
which is common to the entire sample: this assures $\langle L \rangle$ to be obtained from strictly comparable properties of each GRB.
Moreover, no assumption is made on the functional form of the flare temporal profile. 
Finally, since flares are detected as sharp features \emph{superimposed} to the X-ray afterglow, it is of fundamental
importance to understand the associated observational biasses: we compute the temporal evolution of the flare detection threshold
for each GRB, and quantitatively discuss the influence of the flare detection threshold on the obtained $\langle L \rangle$ 
temporal evolution.
  
This work is organised as follows: the sample selection and data reduction is detailed in Sect. \ref{Sec:sample}  while the
computation of the flare and underlying continuum components of each event is presented in Sect.  \ref{Sec:cont}. 
Section \ref{Sec:simulations} describes the method which leads to the definition of a flare detection threshold of each GRB.
A luminosity-time plane avoidance region is presented in Sect.  \ref{Sec:lumtimeplane}. Results are 
discussed in Sect. \ref{Sec:discussion}: possible physical interpretations in the context of both accretion and magnetar 
models of the central GRB engine are critically analysed in Sect. \ref{SubSubSec:slopeaccretion} and 
\ref{SubSubSec:slopemagnetar}. Constraints to physical scenarios able to account for the erratic behaviour of the flare 
emission are derived in Sect. \ref{SubSec:var}.  Conclusions are drawn in Sec. \ref{Sec:summary}.

The phenomenology of the bursts is presented in the rest frame unless otherwise stated. Uncertainties
are quoted at the 68\% confidence level (c.l.): a warning is added if it is not the case. Standard cosmological 
quantities have been adopted: $H_{0}=70\,\rm{Km\,s^{-1}Mpc^{-1}}$, $\Omega_{\rm{\Lambda}}=0.7$, 
$\Omega_{\rm{M}}=0.3$.
\section{Sample selection and data reduction}
\label{Sec:sample}
\begin{table}
\caption{Sample of 44 GRBs observed in the period
April 2005 - February 2010 by \emph{Swift}-XRT with X-ray flaring activity and measured redshift. }
\begin{center}
\begin{tabular}{llll}
\hline
GRB&GRB&GRB&GRB\\
\hline
050730&  060510B& 060926&  080906\\
050814&  060512&   070318&  080928\\
050820A&060526&   070721B&081008\\
050904&  060604&   071031&  081028\\
050908&  060607A&071112C& 090417B\\
051026B&060707&  071122&   090423\\
051022&  060714&  080210&  090516\\
060115&  060729&  080310&090715B\\
060124&  060814&  080607&090809\\
060210&  060904B&080805&090812\\
060418&  060906&  080810&091029\\
\hline
\end{tabular}
\end{center}
\label{Tab:sample}
\end{table}

We select the GRBs which visually show flaring activity superimposed  to the smooth X-ray afterglow, good data 
quality  and redshift, detected by the \emph{Swift} X-ray telescope (XRT, \citealt{Burrows05})  from the beginning of the mission up to the end of February, 2010. The sample
comprises 44 GRBs (Table \ref{Tab:sample}), with redshift ranging from 0.3 to 6.3\footnote{GRB\,090423 with z=8.3 was not
included in spite of the clear X-ray flaring activity to have a wider sample common rest frame X-ray energy band.}.
XRT data have been processed with the \textsc{heasoft} package v. 6.6.1 and corresponding calibration files: standard filtering 
and screening criteria have been applied. 
Piled-up Window Timing (WT) data have been corrected following the prescriptions by
\cite{Romano06}, while piled-up Photon Counting (PC) data have been extracted from an annular region whose inner radius
has been derived comparing the observed to the nominal point spread function (PSF, \citealt{Moretti05}; \citealt{Vaughan06}).
The background is estimated from a source-free portion of the sky and then subtracted. The 0.3-10 keV background subtracted,
PSF and vignetting corrected light-curve of each GRB has been re-binned so as to assure a minimum signal-to-noise (SN) equal to 4.
The count-rate light-curves are calibrated into luminosity light-curves using a time dependent count-to-flux conversion factor
as described in \cite{Margutti10a}. This procedure produces luminosity curves where the possible spectral evolution of the source 
is properly taken into account. Each 0.3-10 keV XRT light-curve is calibrated in the common rest frame energy band defined by the 
redshift distribution of the sample which turns out to be 2.2-14.4 keV. This allows us to make a direct comparison between the X-ray 
afterglows of different bursts while avoiding extrapolation of the signal to an unobserved energy band\footnote{An exception is
GRB\,090417B with z=0.345: this burst was added later to the sample and required the extrapolation of the observed flux from
2.2-13.5 keV rest frame to 2.2-14.4 keV rest frame.}.
\section{Continuum and excess estimation}
\label{Sec:cont}

\begin{figure}
\vskip -0.0 true cm
\centering
    \includegraphics[scale=0.5]{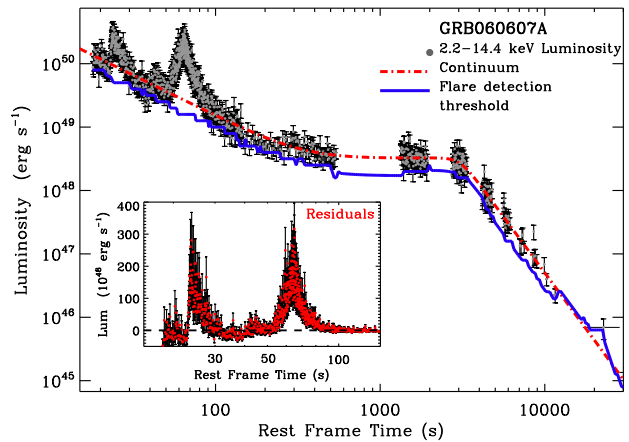}
      \caption{X-ray afterglow of GRB\,060607A in the rest frame energy band 2.2-14.4 keV. Red dot-dashed line: best estimate of the
      continuum underlying the flaring emission obtained as detailed in Sec. \ref{Sec:cont}. Blue solid line: 2$\sigma$ threshold of 
      flare detection (i.e. the minimum peak luminosity that a flare must have to be detected as statistically significant fluctuation 
      \emph{superimposed}  to the X-ray continuum), calculated as described in Sec. \ref{Sec:simulations}. In this case, for $t 
	\lesssim 1000$ s, the threshold is found to be $\sim 50\%$ the continuum.      
      \emph{Inset:} residuals with respect to the continuum.}
\label{Fig:example}
\end{figure}

\begin{figure}
\vskip -0.0 true cm
\centering
    \includegraphics[scale=0.5]{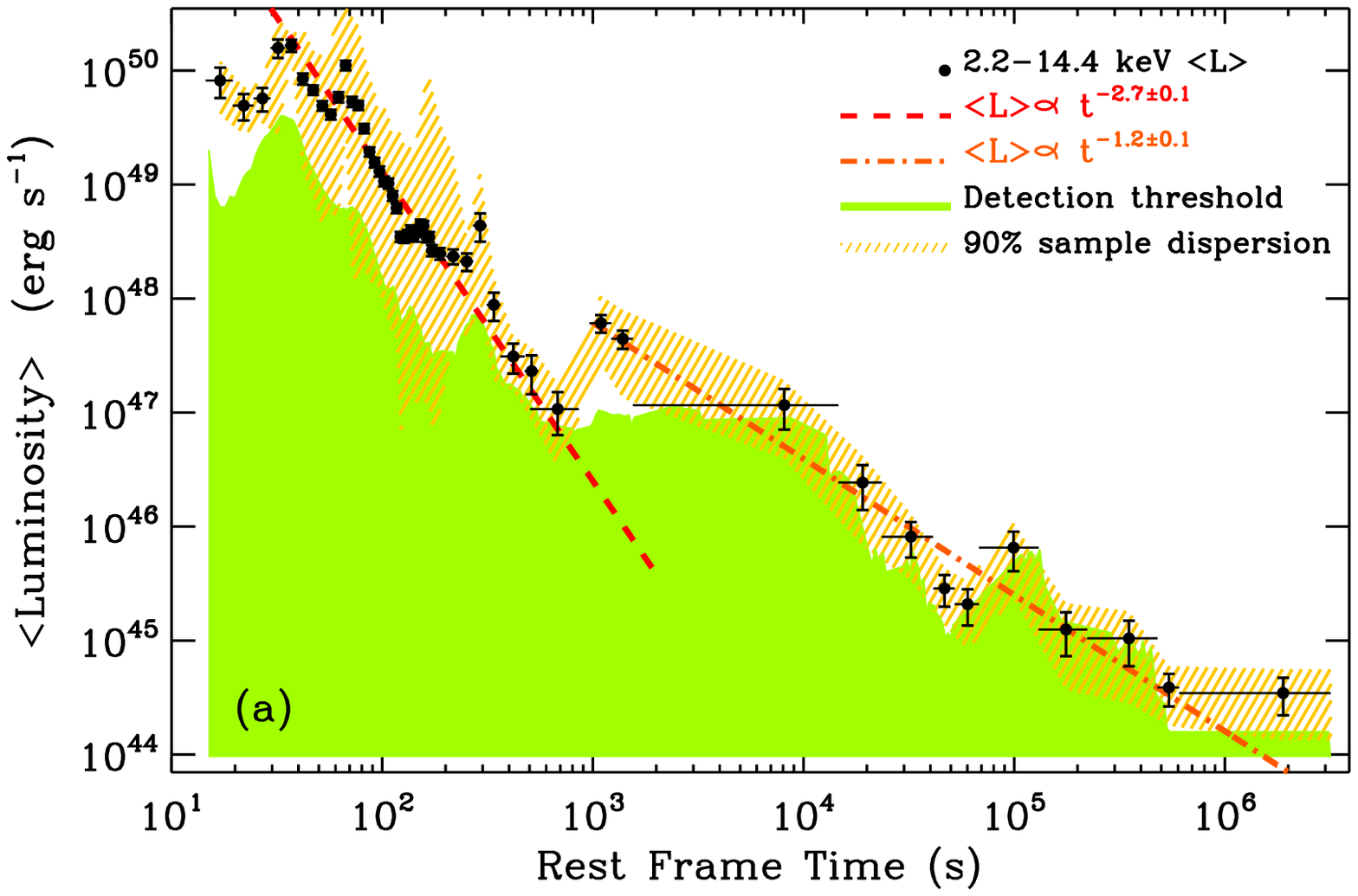}
    \includegraphics[scale=0.5]{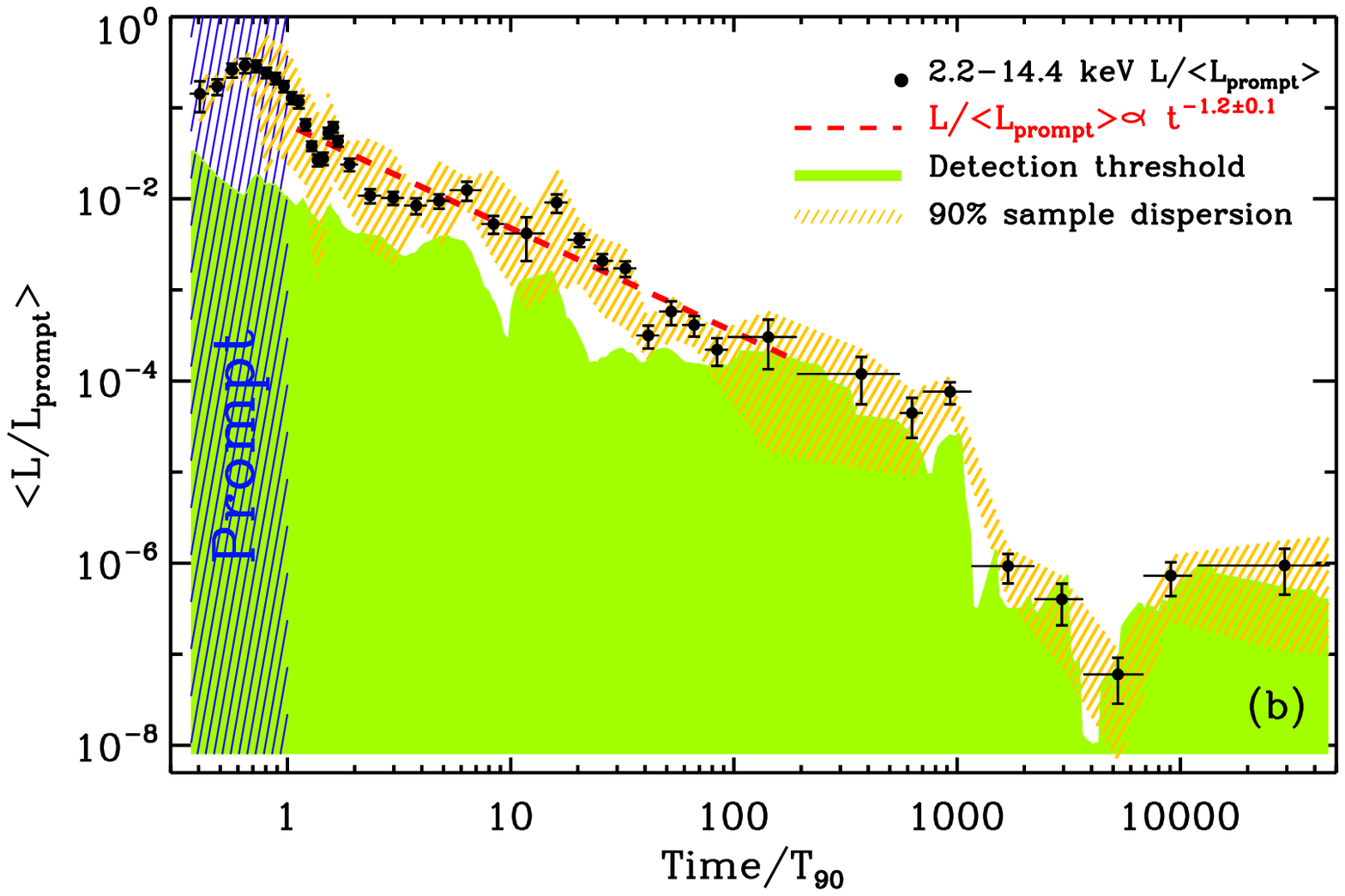}
      \caption{Panel \emph{(a)}: average flaring component luminosity curve of the 44 GRBs computed in the  common rest 
      frame 2.2-14.4 keV energy band. 
      In the time interval $t<1000$ s ($t> 1000$ s) the best fit reads $\langle L/\rm{erg\, s^{-1}} \rangle=10^{54.5\pm0.1} (t/\rm{s})^{-2.7\pm0.1}$
      ($\langle L/\rm{erg\, s^{-1}}\rangle=10^{51.4\pm0.1} (t/\rm{s})^{-1.2\pm0.1}$). Panel \emph{(b)}: average re-normalised luminosity 
      curve: the luminosity has been re-scaled for the average prompt 15-350 keV luminosity observed during the $T_{90}$. The x-axis
      is expressed in $T_{90}$ units. In the time interval $1\lesssim t^{*}\lesssim 100$ the re-normalised flare luminosity follows 
      a power-law decay with best fitting model: $\langle L/L_{\rm{prompt}} \rangle =10^{1.1\pm0.1}(t^{*})^{-1.2\pm0.1}$ where $t^{*}\equiv t/T_{90}$.}
\label{Fig:avelum}
\end{figure}

We produced a software to automatically identify the smooth continuum underlying the X-ray afterglow of GRBs with superimposed
flaring activity. 
The procedure is based on the $\chi^2$ statistics and can be described as a two-step process:
\begin{enumerate}
\item Identify the continuum afterglow component underlying the flaring activity;
\item Calculate the flaring contribution as excess with respect to the continuum.
\end{enumerate}
A first blind fit of the entire X-ray 
afterglow (continuum plus flares) is done in log-log units using a power-law, smoothly joint broken power-law or double broken 
power-law models.  If the P-value  (probability of obtaining a result at least as extreme as the one that is actually observed) 
associated to this fit is lower than 5\%, the data point with the largest \emph{positive} residual is removed from the light-curve and 
a new fit is performed. This process is repeated until a P-value $>5\%$ is obtained. 
The F-test is used to choose between the different nested models when necessary. The best fitting model satisfying the 
P-value condition is identified with the underlying continuum associated to a  particular GRB X-ray afterglow (red dot-dashed line in Fig. 
\ref{Fig:example} for the case of GRB\,060607A) and is subtracted from
the original light-curve. 
The resulting residuals and respective errors (both the statistical uncertainty 
associated to the original light-curve bins and the one coming from the continuum estimation\footnote{We take into consideration 
the complete information coming from the covariance matrix of the fit.} are properly taken into account and propagated) constitute the candidate
X-ray flaring component associated to a GRB: this is shown in the inset of Fig. \ref{Fig:example} for GRB\,060607A. 
Note that no particular flare functional shape is assumed in this analysis. This method allows us to account 
for small variations superimposed to the continuum: in C10 flares were instead visually identified and then fitted with a specific profile.

The average GRB 2.2-14.4 keV rest frame flaring component for any time interval $t_i-t_f$ is computed as:
\begin{equation}
	\langle L(t)\rangle^{2.2-14.4\,\rm{keV}}=10^{\frac{1}{N}\sum_{i=1}^{N}Log(L_{i})}	
	\label{Eq:avelum}
\end{equation}
where N is the number of GRB displaying a positive flaring component at a minimum 2$\sigma$ significance during $t_i-t_f$;
$L_{i}$ is the 2.2-14.4 keV luminosity of the flaring component of the $i^{th}$ burst evaluated at at  $t=(t_i+t_f)/2$.
Linear interpolation is used when necessary.
The uncertainty affecting $\langle L \rangle$ is found through standard propagation of the uncertainties affecting each flaring component,
as determined in the previous paragraph.
The typical uncertainty on each GRB flaring component data point is of the order of $30\%$ the value of the GRB afterglow
light-curve \emph{before} the flare subtraction. An example is shown in  Fig. \ref{Fig:example}, inset.
The use of the \emph{logarithmic} mean in Eq. \ref{Eq:avelum} prevents $\langle L \rangle$ from being dominated 
by a single bright event (the use of the median has 
been proven to lead to very similar results); instead, the linear mean of the excesses leads to $\langle L \rangle$ values biassed 
towards the bright end of the luminosity distribution of the excesses at any time $t$, and is therefore discarded. 
GRB\,050904 shows an extended flaring activity between 500 s and 3.1 ks: given the high luminosity of the flaring component
and the peculiarity of a flaring emission covering the entire plateau phase duration, GRB\,050904 is not included in the calculation
above. This guarantees that  $\langle L \rangle$ is not biassed by the contribution of a single peculiar burst: however, the inclusion of this event
would not change the main conclusions of this work. The flare luminosity curve of GRB\,050904 is investigated and discussed 
in Sec. \ref{Sec:discussion}. 
The result of the application of Eq. \ref{Eq:avelum} 
is shown in Fig. \ref{Fig:avelum}, panel \emph{(a)}: in the time interval $15-30$ s the average light-curve of the flaring component is likely to 
track the prompt gamma-ray emission, while  for $30<t<1000$ s the average light-curve is well represented by a
power-law with best-fitting index $\alpha_1=-2.7\pm0.1$. For $t>1000$ s the best-fitting power-law index reads $\alpha_2=-1.2\pm0.1$. 
We note that a simple power-law modelling of the entire $15\,\rm{s}<t<10^6\,\rm{s}$ temporal window yields a statistically
unacceptable fit
($\chi^2/\rm{dof}=1271.4/50$, P-value$\ll 10^{-10}$) that systematically overestimates the computed average flare luminosity for $t>1$ ks.
The two data points at $t\sim10^3$ s largely above the detection threshold (green thick area, see Sec. \ref{Sec:simulations} for details) are 
mainly due to the flaring contribution of GRB\,081028, GRB\,090809 and GRB\,090417B: these 3 events show a large flare at this rest
frame epoch. The exclusion of these 3 bursts from Eq. \ref{Eq:avelum} would bring the two data points slightly above the threshold.
The shown error bars account for the $1\sigma$ uncertainty affecting the average luminosity value: however, 
the dispersion of the sample luminosity values at any given time $t$ could be remarkably different. For this reason a  
yellow line-fillled area marking the 90\% sample dispersion has been added to  Fig. \ref{Fig:avelum}, both panels.

The GRB prompt $\gamma$-ray emission is known to span different orders of magnitude in luminosity.
We correct for the different prompt luminosity characterising different events re-normalising the flaring component by the 
average 15-350 keV luminosity observed during the BAT $T_{90}$ (interval of time of emission of 90\% of the BAT fluence). 
The re-normalised flaring luminosity $\langle L^{*}\rangle \equiv \langle L/L_{\rm{prompt}} \rangle$ in the time interval $t_{i}^{*}-t_{f}^{*}$ is computed as:
\begin{equation}
	\label{Eq:avelumren}
	\langle L^{*}(t^{*})\rangle ^{2.2-14.4\,\rm{keV}}=10^{\frac{1}{N}\sum_{i=1}^{N}Log(L_{i}^{*})}		
\end{equation}
where $L_{i}^{*}\equiv L_{i}/\langle L_{i,\gamma} \rangle$, being $\langle L_{i,\gamma} \rangle$ the average 15-350 keV prompt luminosity during the 
$T_{90}$ of the $i^{th}$ burst. For each burst, the time $t^{*}$ is expressed in $T_{90}$ units\footnote{GRB\,051022
was detected by HETE and consequently has no measured BAT $T_{90}$: for this reason it is not included in the calculation.
GRB\,050904 is also excluded because of its peculiarity (see Sect. \ref{Sec:discussion}).}.
Both the BAT $T_{90}$ and the fluence information are retrieved from the burst BAT-refined circulars\footnote{
For GRB\,060424 we refer to \cite{Romano06}.}.  
Figure \ref{Fig:avelum}, panel \emph{(b)}, shows the resulting 
re-normalised average flaring luminosity curve: for $t^{*}\lesssim1$ the XRT caught the X-ray counterpart of the prompt $\gamma$-ray 
emission; in the time interval $1\lesssim t^{*}\lesssim100$ the re-normalised flaring activity follows a power-law decay with best-fitting
index $\alpha=-1.2\pm0.1$. 
\section{Flare detection threshold}
\label{Sec:simulations}

\begin{figure*}
\vskip -0.0 true cm
\centering
    \includegraphics[scale=0.85]{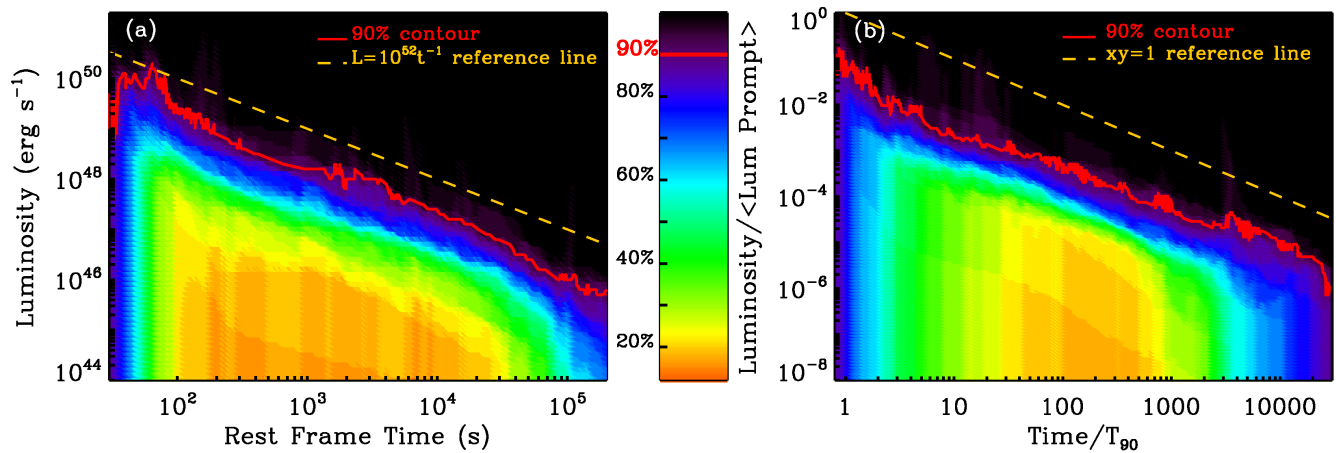}
      \caption{Panel \emph{(a):} X-ray flaring activity contour plots in the luminosity-time plane. The area is color coded 
	as defined in the middle panel. The luminosity is 
      computed in the rest frame 2.2-14.4 keV energy band. The flaring activity of 90\% of 
      the GRBs in our sample lies below the red solid line. Yellow dashed line: power-law with index -1 plotted for comparison.
      Panel \emph{(b)}: the luminosity is expressed in units of the average 15-350 keV prompt luminosity during the  BAT $T_{90}$;
      times are in $T_{90}$ units while the time is expressed in units of BAT $T_{90}$.}
\label{Fig:lumtimeplane}
\end{figure*}

Flares are always detected as emission components \emph{superimposed} on top of the overall afterglow decay: this means that
whatever the origin of the continuum emission is, the sensitivity to temporal fluctuations is degraded  as the flux level -and
consequently the statistics- of the underlying continuum decays with time (see e.g. \citealt{Morris08}). 
This naturally requires flares at late times to be  
longer in duration and/or have a higher flux contrast with respect to the continuum in order to be detected as statistically
significant fluctuations. It is therefore of primary importance to understand at which level the evolution with time of the 
average flaring component portrayed in Fig. \ref{Fig:avelum} is a by-product of the decay of the underlying X-ray afterglow  
which acts as a time-variable detection threshold.

To this end a set of simulations is run for each GRB of our sample: the aim is to determine the flux and luminosity detection 
threshold of the flaring component at any time $t$ given the observed X-ray afterglow statistics.
The continuum contribution to each afterglow and associated uncertainty have been calculated in Sect. \ref{Sec:cont}: 
we evaluate the continuum at the original light-curve central time bins and combine the systematic uncertainty 
coming from the continuum evaluation procedure with a statistical uncertainty equal to the relative error on the original
luminosity bins of that particular burst. A grid of 545 flare peak time values $t_{pk}$ equally spaced in logarithmic units between
1 s and $10^7$  s is generated together with a grid of 171 flare peak luminosity values in the interval $10^{40}-10^{57}\,\rm{erg\,s^{-1}}$.
For each $t_{pk}$, we generate 171 flares with growing peak luminosity values: a  \cite{Norris05} profile specified in
terms of amplitude $A$, width $w$ and peak time $t_{pk}$ is assumed. The Norris 2005 profile is defined as $f\equiv f(A,\tau_1,\tau_2,t_s)$
where $\tau_1$ and $ \tau_2$ are two flare shape parameters related to the rise and decay phases, respectively, while $t_s$ defines the
pulse starting time. We take advantage of previous studies on the flare phenomenology to reduce the flare parameter space and write 
$\tau_1\sim4w/3 $; $\tau_2\sim w/3$; $t_s\sim t_{pk}-2w/3$; $w\sim10^{-0.94}t_{pk}^{(1.12\pm0.18)}$ (see C10 and 
M10 for details).  
The signal of the fake flare is then integrated in the original light-curve bins 
and a statistical error is assigned equal to the relative uncertainty affecting the original luminosity curve at the same time.
Finally, the flare contribution is
added to the continuum and the uncertainties propagated into a final fake GRB afterglow. To simulate the entire procedure, the fake 
afterglow is then processed by the automatic excess estimation routine described in the previous section: a new continuum
is determined together with the associated flaring component, as before. If the newly determined flaring component 
contains a \emph{positive} fluctuation with a minimum $2\sigma$ significance, the peak luminosity of the fake flare is recorded 
as $2\sigma$ flare detection threshold at $t_{pk}$ and the following peak time value is considered. 
The entire procedure is repeated for each $t_{pk}$,  every time starting from the minimum peak luminosity value of the grid. 
The result is shown in Fig. \ref{Fig:example} for GRB\,060607A taken as an example (blue solid line): due to the degrading quality 
of the signal, a growing flare-to-continuum flux contrast is required for the flare to be detected at later times.
  
The $2\sigma$ detection luminosity curve of the different GRBs are then combined to produce the average detection 
threshold $\langle L_{th}(t) \rangle$ in the rest frame band 2.2-14.4 keV. In strict analogy with Eq. \ref{Eq:avelum}, $\langle L_{th}(t) \rangle$ is
defined as:
\begin{equation}
	\langle L_{th}(t)\rangle ^{2.2-14.4\,\rm{keV}}=10^{\frac{1}{N}\sum_{i=1}^{N}Log(L_{i,th})}	
\end{equation}
being $N$ the number of GRBs displaying a significant positive flaring component in the time period $t_i-t_f$;
$L_{i,th}$ is the luminosity of the $2\sigma$ detection threshold of the $i^{th}$ GRB at time $t=(t_i+t_f)/2$.
The re-normalised average $2\sigma$ threshold $\langle L^{*}_{th}(t^{*}) \rangle$ is computed  as:
\begin{equation}
	\langle L^{*}(t^{*})\rangle^{2.2-14.4\,\rm{keV}}=10^{\frac{1}{N}\sum_{i=1}^{N}Log(L_{i,th}^{*})}			
\end{equation}
where $L_{i,th}^{*}\equiv L_{i,th}/\langle L_{i,\gamma} \rangle$. As before, $N$ collects the GRBs with significant excesses in 
$t_{i}^{*}-t_{f}^{*}$. 

The result is portrayed in Fig. \ref{Fig:avelum}: in both panels a thick green area marks the $2\sigma$
detection threshold region. From panel $(a)$ it is clear that the $-1.2$ slope is mainly due to the \emph{selection} of bright fluctuations 
able to overshine the threshold. The fact that $\langle L(t) \rangle $ tracks the $\langle L_{th}(t)\rangle$ temporal behaviour suggests that the 
sample of significant excesses detected in the time period 2-3000 ks are not representative of the real population of flares
at those times. The bright end of the flare luminosity distribution has been sampled and a biassed average flare luminosity
has been re-constructed. Notably, the decay of the detection
threshold at $t\sim10^4$ s allows us to appreciate a steeper decay ($\alpha\sim -2.5$) of $3\sigma$
significant positive fluctuations extending to $t\sim6\times 10^4$ s. No more than $2\sigma$ significant flaring activity 
can be quoted for $(1<t<6)\times 10^5$ s due to the flatter decay (and actually enhancement) of $\langle L_{th}(t) \rangle$.
The unbiassed average flare luminosity curve is therefore likely to be steeper than $-1.2$ at late times. At early times
the situation is different: for $t\lesssim 300$ s the $\langle L(t) \rangle \propto t^{-2.7}$ decay is not affected by threshold effects;
the extension of the  same temporal law up to $t\sim1000$ s suggests that $\langle L_{th}(t) \rangle$ is likely to play a minor
role in the time interval $(0.3<t<1)$ ks, as well. Finally, Fig. \ref{Fig:avelum}, panel \emph{(b)}  shows that the 
re-normalised average flare luminosity tracks the threshold  starting from $t^{*}\sim100$. $\langle L^{*}(t^{*}) \rangle$ is consistent
with the $-1.2$ power-law decay up to $t^{*}\sim800$. For $t^{*}\geq1000$ the re-normalised activity significance
is $\lesssim 2\sigma$.

\section{Luminosity-time plane: the flare avoidance region}
\label{Sec:lumtimeplane}
The $2\sigma$ detection threshold luminosity curves calculated in Sec.  \ref{Sec:simulations} define for each GRB,
for each time $t$, the minimum peak luminosity which a flare must have to be detected as statistically significant positive 
fluctuation above the X-ray afterglow (see Fig. \ref{Fig:example}, blue solid line). 
We combine this information with the residual luminosity curves 
derived in Sec. \ref{Sec:cont} to draw an upper limit to the X-ray flaring luminosity associated to each  afterglow at any time $t$, 
as follows: for each event we substitute the luminosity value of non significant fluctuations 
with the respective $2\sigma$  upper limits from the corresponding threshold 
luminosity curve, keeping the significant fluctuations unchanged.
In this way we account for the presence of 
\emph{undetectable} flaring activity. 
The resulting curve divides the luminosity-time plane into two halves: the real flaring luminosity curve is expected to lie 
below this line at a minimum 2$\sigma$ c.l.: we refer to this region as the X-ray flare-area associated to a particular GRB.
This procedure is repeated for each burst of the sample for both luminosity and re-normalised luminosity units.
The different flare-areas are then superimposed and a luminosity-time plane contour plot is created (Fig. \ref{Fig:lumtimeplane}):
the different colours refer to the different number of GRB flare-areas superimposed  on a particular $(t_i,L_i)$ pixel.
In both panels the flaring luminosity of the 90\% of the GRBs of our sample lies below the red solid line\footnote{In the time 
interval $60\,\rm{s}<t<2\times 10^{5}\,\rm{s}$ the 90\% contour of Fig. \ref{Fig:lumtimeplane}, panel \emph{(a)} is well approximated by: 
$5.6\times 10^{48}[(\frac{t}{200})^{-5.1}+(\frac{t}{200})^{-1.3}]^{0.5}[1+(\frac{t}{3162})^{2.2}]^{-0.3}$. For panel \emph{(b)}, the
approximation reads: $4.5\times 10^{-3}[(\frac{t}{5})^{-5.5}+(\frac{t}{5})^{-2.6}]^{0.3}$}. We refer to the area above the red line as 
the luminosity-time plane flare avoidance region. Note that a direct implication of Fig. \ref{Fig:lumtimeplane}, panel \emph{(a)}
is that in the time period $60\,\rm{s}\lesssim t\lesssim 400\,\rm{s}$ the flare luminosity function decay is steeper than 
$\propto t^{\sim-1.8}$ at 90\% confidence. 
\section{Discussion}
\label{Sec:discussion}
\begin{figure}
\vskip -0.0 true cm
\centering
    \includegraphics[scale=0.45]{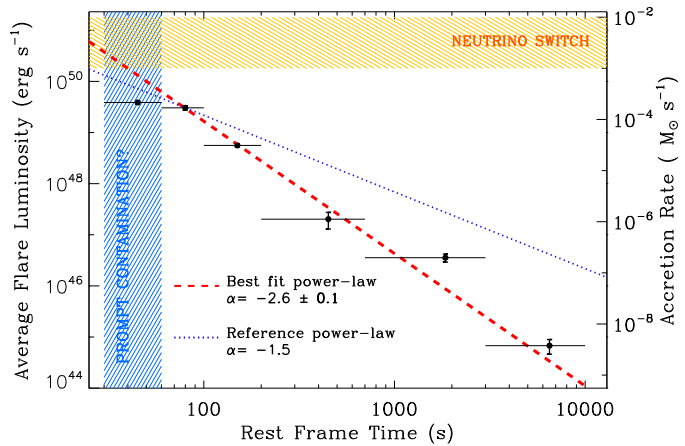}
      \caption{Average flare luminosity curve for the enlarged sample of 55 flares belonging to 29 different GRBs with known red-shift
	calculated following the prescriptions by Lazzati et al., 2008. 
      1$\sigma$ uncertainties affecting the \emph{average} value have been shown.
	Red-dashed line: best-fit power-law with index $\alpha=-2.6\pm0.1$. 
	As in Lazzati et al. 2008 the first data point has been excluded from the fit.  Blue-dotted line: reference $\alpha=-1.5$ power-law; the 
	normalisation has been chosen so as to intercept the second point shown. Corresponding accretion-rate units are provided on the right 
	y-axis: $L_{\rm{iso}}\Omega_{\rm{beam}}=\eta_{\rm{acc}}\dot Mc^2$, with $\Omega_{\rm{beam}}=0.01$, $\eta_{\rm{acc}}=0.001$. The 
	orange hatched area approximately marks the boundary between the NDAF (neutrino-dominated accretion flow) and ADAF 
	(advection-dominated accretion flow) regimes when the neutrino cooling switches off (see e.g. 
	Narayan et al. 2001; Chen \& Beloborodov 2007).} 
\label{Fig:lazzatiplot}
\end{figure}

In the previous sections we showed that:
\begin{itemize}
	\item For $15\,\rm{s}<t\lesssim30\,\rm{s}$ the average flare luminosity function is likely to track the prompt $\gamma$-ray emission;
	\item In the time interval $30\,\rm{s}\lesssim t \lesssim 1000\,\rm{s}$ we find $\langle L \rangle \propto t^{-2.7\pm0.1}$.  
	No threshold-related argument can be invoked to explain this scaling. 
	\item For $t\gtrsim 1000$ s $\langle L \rangle $ closely follows the temporal behaviour of the threshold of detection: 
	$\langle L \rangle \propto t^{-1.2\pm0.1}$.
	For this reason we believe the un-biassed average flare luminosity function at this epoch to be steeper. 
\end{itemize}

Notably, a similar $L_{\rm{pk}}-t_{\rm{pk}}$ scaling was obtained by \cite{Chincarini10} for $t<1000$ s
once the intrinsic scatter of the relation was properly accounted for: $L_{\rm{pk}}\propto t_{\rm{pk}}^{-2.7\pm0.5}$, where $L_{\rm{pk}}$
and $t_{\rm{pk}}$ are the flare peak luminosity and flare peak time as obtained from a \cite{Norris05} profile fit. 

In the following we therefore focus our attention on the $30\,\rm{s}< t < 1000\,\rm{s}$
interval of time, where the flare luminosity function $\langle L \rangle \propto t^{-2.7}$. In Sect. \ref{SubSec:tslope} we first
investigate the source of the discrepancy between our results and those reported in L08. Possible physical interpretations
in the context of both accretion and magnetar models of the central GRB engine are discussed in Sect. \ref{SubSec:tslope}
and  \ref{SubSec:var}. 
\subsection{The $L \propto t^{-2.7} $ regime}
\label{SubSec:tslope}
The GRB average X-ray flare luminosity function was first analysed by L08 starting from a sample of 24 flares coming from
9 long GRBs and 1 short GRB. The mean luminosity was found to decline with a much shallower power-law in time: 
$\langle L \rangle _{\rm{L08}}\propto t^{-1.5\pm0.2}$. This scaling was obtained averaging the contribution of single flares of assumed square shape over time-scales 
longer than the flare duration. We investigate the source of the discrepancy ($\langle L \rangle _{\rm{L08}}
\propto t^{-1.5}$ vs. $\langle L \rangle \propto t^{-2.7}$ of our work) below.
We follow the prescriptions of L08 and apply their method to the sample of early time flares of C10
supplemented by late time flares with known redshift: the enlarged sample includes 55 flares detected in 29 GRBs.  We do not assume a flare square 
shape and used instead the best fitting parameters of a \cite{Norris05} profile: however, the choice of a particular flare profile is unlikely 
to bias the final result, as already noted by L08. The interval of times of integration of the mean luminosity function 
have been estimated from their Fig. 1. Figure \ref{Fig:lazzatiplot} shows the final result: the average flare luminosity function is best
modelled by a power-law with best fitting index $\alpha=-2.6\pm0.1$. The shallower $t^{-1.5}$ scaling is also represented for comparison.
Errors are computed following the standard theory of error propagation. This result confirms the findings of Sect. \ref{Sec:cont} with a 
completely independent method. The reason for the discrepancy lies elsewhere.

We further check the consistency between the L08 method and ours, measuring the flaring activity evolution of GRB\,050904. 
In this case we find $L\propto t^{-0.95\pm0.05}$ which is consistent with $L\propto t^{-1.0\pm0.3}$ quoted by L08. The flaring
evolution of GRB\,050904 is much flatter than the average behaviour $\propto t^{-2.7}$.  Prompted by this result we investigate
the temporal evolution of the flaring component in the subsample of GRBs of Table \ref{Tab:sample} showing multiple
episodes of flaring emission. The subsample contains 10 elements (black dots in Fig. \ref{Fig:slopetoslope}): the mean flare function
decay index is $-1.8$. This suggests that multiple flare GRBs have a flatter than average flare luminosity function. 
Interestingly, this subsample also shows a flatter than average steep decay: the median temporal decay index is found to be
$\alpha_{\rm{steep}}=1.7$ to be compared to $\alpha_{\rm{steep}}=2.5$ of the remaining 34 GRBs of the original sample. This 
topic is further explored in  Sec. \ref{SubSec:var}.
The average number of flares per GRB in the L08 sample is $\langle n_{f} \rangle=2.4$ while our enlarged sample 
contains a wider population of single-flare GRBs: $\langle n_{f}\rangle =1.9$. In particular, $60\%$ of the GRBs of L08  contains more 
than one flare, while only $25\%$ of the enlarged sample shows more than one episode of activity. The discussion above leads
to the conclusion that the choice of a sample biassed towards multiple-flare GRBs (which was the wider sample of flares with 
known redshift available at the time of writing), led L08 to determine a flatter mean luminosity temporal scaling. 
In agreement with this picture, in the two GRBs with multiple flares for which L08 determined the flare function evolution
they obtained $\alpha=-1.0\pm0.4$ (GRB\,050803) and $\alpha=-1.0\pm0.3$ (GRB\,050904): both values are flatter 
-even if consistent- than the average $-1.5\pm0.2$ decay index derived from their entire sample.

While we benefit from a higher statistics which makes our sample more representative of the entire bursts population,
the same choice of GRBs with easily recognisable flares in their afterglows is possibly introducing a bias which is difficult
to quantify. The analysis of the flare avoidance region of Sect. \ref{Sec:lumtimeplane}, which properly accounts for the
fraction of undetectable flares is the safest approach in this respect: this analysis indicates the average flare luminosity
decay to be steeper than $\propto t^{-1.8}$ at the $90\%$ c.l. for $t\lesssim 400$ s. 

We finally investigate if the different prompt luminosity (and consequently afterglow luminosity) of the 44 GRBs of Table
\ref{Tab:sample}, could bias the flare $\langle L \rangle$ temporal scaling. Re-normalising the flaring component of each GRB by
the average prompt luminosity during the burst $T_{90}$ duration\footnote{Note that this is different from what displayed in Fig. 
\ref{Fig:avelum}, where the x-axis is in $T_{90}$ units.}  the $\langle L/L_{\rm{prompt}} \rangle$ is best fitted by a power-law decay with index
$\alpha\sim-2.5$ for $t<1000$ s.

The $\propto t^{\sim -2.7}$ behaviour of the average flare luminosity has been obtained in 4 completely independent ways:
\begin{enumerate}
	\item Equation \ref{Eq:avelum} applied to 44 GRBs;
	\item Equation \ref{Eq:avelum} with re-normalised flaring contribution; 
	\item $L_{\rm{pk}}$ vs. $t_{\rm{pk}}$ relation from C10;
	\item Application of the L08 method to the enlarged sample of 29 long GRBs  with known redshift fitted with a \cite{Norris05} profile.
	\end{enumerate}
It is important to underline that the $\propto t^{-2.7}$ behaviour represents the \emph{average} scaling of the GRB flaring activity:
the decay of the flaring component in a particular GRB can considerably differ from the average relation, as shown by Fig. 
\ref{Fig:slopetoslope} (see also Fig. \ref{Fig:avelum}, upper panel, where the yellow hatched area marks the 90\% sample dispersion).
Any model aiming at explaining the flare phenomenology is required to provide an explanation for the observed scatter, as well. 
Different physical interpretations are discussed below.
\subsubsection{Accretion-powered flares scenario}
\label{SubSubSec:slopeaccretion}

In the case of a thin disk\footnote{Note however that at the high mass accretion rate expected in the collapsar model a geometrically \emph{thick} disk
is expected (see e.g. Kumar et al. 2008a,b; \citealt{Narayan01}).} 
with constant viscosity for $t\gg t_{\alpha_{\nu}}$ (where $t_{\alpha_{\nu}}$ is the viscous time-scale)
the accretion rate approaches the asymptotic regime $\dot m \propto t^{-1.25}$ (\citealt{Franck85}). A similar
slope is obtained by \cite{Cannizzo90} performing numerical simulations using the $\alpha$-viscosity prescription 
of \citealt{Shakura73}: $\dot m\propto t^{-1.2}$. Recently \cite{Metzger08} computed  the viscous evolution 
of an efficiently $\nu-$cooled, isolated ring of material under similar assumptions as the study of \cite{Cannizzo90}: they report 
a self-similar behaviour $\dot m \propto t^{-4/3}$. The $t^{-2.7\pm0.1}$ average flare luminosity power-law scaling obtained 
in this work is much steeper than the theoretical expectations listed above: an ad hoc non-linear relation between the luminosity
and the outflow accretion rate, with the efficiency of conversion of mass accretion into luminosity decreasing as
$L/\dot m \propto t^{-1.5}$, is required to preserve the thin disk draining accretion scenario. \cite{Lee09} demonstrated that
the presence of powerful winds driven by the recombination of $\alpha-$particles into nucleons and launched from the disk surface 
could lead to a substantial deviation from the $t^{-4/3}$ regime, with $\dot m$ entering an exponential-like decay. This regime is unlikely to 
extend up to $t\sim 100-1000$ s (see \citealt{Lee09}, their Fig. 1). 
Alternatively, a temporal evolution of the flare opening angle causing later flares to be less beamed than earlier ones would restore 
the $\sim -1.5$ slope\footnote{In the present work each plot refers to the \emph{isotropic} equivalent luminosity of the X-ray flares at any time $t$.}.  
However, most of the beaming evolution has been proven to take place at early times (see e.g. \citealt{Morsony07} and references therein) 
and is consequently unlikely to cause an overestimation of the flare luminosity steepness. 
We therefore consider alternative scenarios. 

Fallback of the stellar envelope that did not reach the escape velocity during the explosion can provide the source of 
accreting material at late times: \cite{Chevalier89} showed that in this case the accretion rate is expected to decline as 
$\dot m \propto t^{-5/3}$ which is too shallow to explain the observed $t^{-2.7\pm0.1}$ flare luminosity scaling.   
The process was later investigated by \cite{MacFadyen01}: these authors simulated the fallback of the stellar envelope 
following the failure of the shockwave to unbind the star and recovered the $\dot m_{\rm{Rmin}}\propto t^{-5/3}$ dependency
(where $\dot m_{\rm{Rmin}}$ is the radial fallback rate through their inner numerical boundary $R_{\rm{min}}=10^9$ cm).
However, departures from the $t^{-5/3}$ scaling are expected if the infalling material passes through an accretion shock
developing at $R<10^9$ cm or/and if the mass fall-back rate decreases suddenly at some time $t$. These possibilities are 
discussed below.

Kumar et al., (2008a,b) suggested that the prompt and X-ray afterglow reflect the modulation in the rate of central accretion 
of a rotating progenitor with a core-envelope structure onto a black hole. A \emph{thick} disk configuration is assumed.
According to their analytical treatment the
condition $\dot m\propto t^{-3}$ can be basically achieved in two ways: first, in the case of rapid accretion, ($t_{\rm{acc}}\ll t$),
$\dot m$ is expected to track the temporal evolution of the mass fall-back rate $\dot m_{\rm{fb}}$. Assuming a 14 $M_{\sun}$
GRB progenitor star (model 16T1 of \citealt{Woosley06}),  Kumar et al., (2008a,b)  expect $\dot m_{\rm{fb}}\propto t^{-3}$
(which directly translate into $\dot m\propto t^{-3}$) when the last $\sim 0.5\,M_{\sun}$ of the star near the surface is accreted. 
A second possibility arises when the stellar collapse leaves behind an accreting disk and no further mass is being added:
$\dot m_{\rm{fb}}=0$. The black hole accretion rate is found to scale as: $\dot m\approx \dot m_{0}t^{-4(s+1)/3}$ (\citealt{Kumar08b},
their Eq. 43)\footnote{This result is valid for $t\gg t_{\rm{acc}}(t_0)$ where $t_{\rm{acc}}$ is the viscous time scale of accretion.
The scaling of the accretion rate with radius has been parametrised as $\dot m \propto (r/r_{\rm{d}})^s$ being $r_{\rm{d}}$ the
accreting disk radius.}. The steepest decline allowed by these conditions is $\dot m\propto t^{-8/3}$, suggestively close  to the average 
flare luminosity $t^{-2.7}$ temporal decay. This regime establishes for $s=1$, when a \emph{convection}-dominated accretion flow
(CDAF, see e.g. \citealt{Narayan01} and references therein) sets in and the accretion rate decreases linearly with disk radius. 
In reality both the steeply falling density profile of the GRB progenitor and the loss of accreting matter in a wind (which is the 
signature of a CDAF or ADAF scenarios), are likely to play a major role in decreasing the fraction of mass which effectively 
makes it to the black hole. These two effect coupled together are able to account for $L_{\rm{jet}}\propto t^{-3}$ or steeper,
for $s\gtrsim 0.5$ (see \citealt{Kumar08b}, their Fig. 4).

This scenario was further explored and confirmed by \cite{Lindner10}. Using axisymmetric hydrodynamical 2D simulations, 
\cite{Lindner10} showed that a \emph{steady} accretion rate  $\dot m \sim 0.1-0.2\, M_{\sun}\,s^{-1}$ giving rise to the prompt $\gamma$-ray 
phase is followed at $t \gtrsim 40 $ s by a sudden and rapid power-law decline of the central accretion rate which is responsible 
for the prompt-to-steep decay X-ray afterglow transition. The  steep decline is triggered by a rapid outward expansion of an 
accretion shock through the material feeding a \emph{convective} thick accretion disk and lasts $\sim 500$ s. During this phase the 
central engine is supposed to be active and the accretion rate is found to decay as $\dot m \propto t^{-2.7}$ (\citealt{Lindner10}).
Again, we find the average flare luminosity function to show a very similar scaling ($\propto t^{-2.7\pm0.1}$ for $t<1000$ s), 
implying a linear relation between the luminosity and accretion rate:  $L\sim\eta_{\rm{acc}}\dot m$.  The efficiency $\eta_{\rm{acc}}$ of mass 
rate to jet luminosity conversion depends on many parameters; among these, the black hole spin $a$ is likely to play a major role.
In Fig. \ref{Fig:lazzatiplot} we use $\eta_{\rm{acc}}=10^{-3}$ which corresponds to $a=0.75$ following the prescriptions of 
\cite{McKinney05}. The main limitation of this study lies in  that \cite{Lindner10} do not simulate the fluid magnetic field
which could alter the $\dot m(t)$ dependence.
\subsubsection{Magnetar models}
\label{SubSubSec:slopemagnetar}
Pulsars arise from core-collapse. A subset of them are inferred to host magnetic fields $B\sim 10^{14}-10^{15}$ G. The rotational 
energy of a rapidly rotating magnetar can be extracted to power a strong relativistic jet (e.g. \citealt{Bucciantini09}) that would give
rise to the prompt emission. The energy release in radiation of this system at late times ($100\,\rm{s}\lesssim t \lesssim 1000$  s)
is subject to great uncertainties (T. Thompson, private communication). However, the output of radiation is likely to be \emph{between}
two limiting cases: dipole radiation and monopole radiation. In the former case $L\propto (1+t/T)^{-2}$ which implies $L\propto t^{-2}$
for $t\gg T$, where $T$ is the characteristic spin-down time-scale (e.g. \citealt{Wheeler00}). In the latter, $L\propto e^{-2t}$
(\citealt{Thompson04}). Based on the braking indices of observed pulsars in the Galaxy, $L\propto t^{-2.3}$ is expected (\citealt{Bucciantini06}). 
Additionally, 
the interaction with the stellar material can be responsible for shallower slopes detected in multiple-flare GRBs (\citealt{Lazzati08}).
\subsection{The variability issue}
\label{SubSec:var}
\begin{figure}
\vskip -0.0 true cm
\centering
    \includegraphics[scale=0.42]{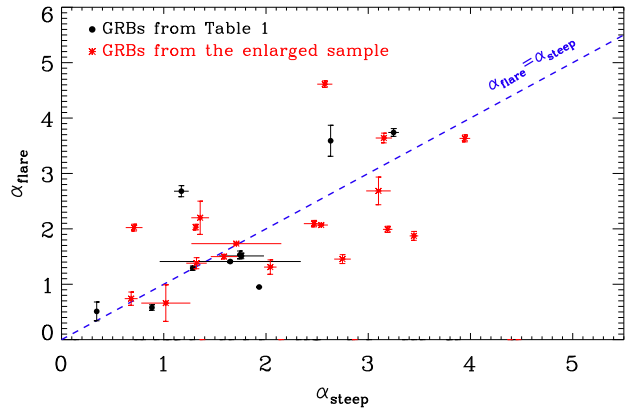}
      \caption{Best fit power-law index of the temporal decay of the X-ray flaring activity vs. continuum (i.e. steep decay) power-law index for the 
      subsample of GRBs listed in Table \ref{Tab:sample}  with extended flaring emission (black dots). GRBs which show multiple
      X-ray flares but otherwise lack of the redshift measurement are indicated with red stars. These events are not part of the main sample
      of Table \ref{Tab:sample}.}
\label{Fig:slopetoslope}
\end{figure}

\begin{figure}
\vskip -0.0 true cm
\centering
    \includegraphics[scale=0.52]{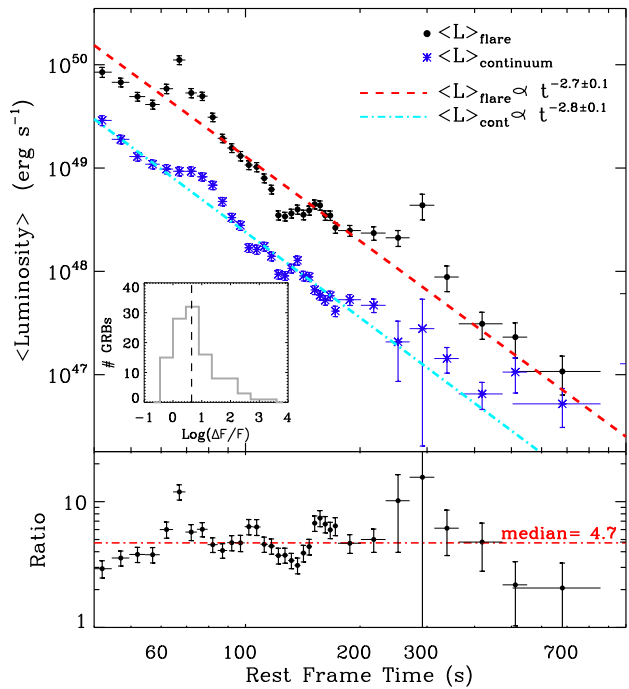}
      \caption{\emph{Upper panel}: average flare luminosity function (black dots) compared to the average luminosity of the continuum underlying the
      flare emission (blue stars) in the time interval $40\,\rm{s}<t<1000\,\rm{s}$. The best fit of both components is also shown. \emph{Inset}:
      flare-to-continuum flux ratio for the sample of early time flares of C10. The dashed line marks the median value of the distribution, which is 
      $\Delta F/F=4.4$. \emph{Lower panel}: 
      flare-to-continuum ratio as a function of time. The median value of the distribution is $\langle L_{\rm{flare}}\rangle/\langle L_{\rm{cont}}\rangle=4.7$. }
\label{Fig:continuum}
\end{figure}

Flares are observed as episodic, large-scale amplitude variations in the light-curves with a typical $\Delta t/t \approx 0.1$ (\citealt{Chincarini10}): 
this automatically raises the question of the physical source of the flare variability. The presence of a correlation between the properties
of the prompt and flare emission within the same GRB could in principle shed light on the mechanism powering the flaring activity. However
we found no correlation between the prompt luminosity and the total energy and luminosity of the flaring 
component; the same is true if one were to consider the 15-150 keV prompt energy; 
no correlation has been found between the prompt luminosity (or energy) and the peak time of the last flare;  
the peak time of the last flare is also not correlated with the prompt $T_{90}$;
the number of prompt pulses is not the key
factor determining the number of flares to appear at $t\gtrsim 50$ s (\citealt{Chincarini07}); finally, a hint for a prompt-fluence vs. flare-fluence
correlation was recently reported by C10: however, the weak correlation is mostly due to the presence of two short GRBs.
The conclusion is that, while flares do share with prompt pulses several key observational properties (a notable example is the lag-luminosity 
relation of M10), at the moment it is not possible to infer the flaring activity of a burst starting from its prompt emission.

Flares are under-represented in simple power law X-ray afterglows (M10). The X-ray afterglow morphology - flaring activity
connection is here further explored starting from the findings of Sec. \ref{SubSec:tslope}: multiple flare GRBs have on average flatter 
flare luminosity functions (i.e. $\propto t^{-\alpha}$ with $\alpha< 2.7$). This, together with the observation that the flare detection threshold
of Fig. \ref{Fig:avelum}, upper panel, closely tracks the temporal decay of the average flaring activity for $t\lesssim 300$ s, brought us to 
consider the possibility of a correlation between the X-ray flare luminosity decay and the underlying continuum emission decay.
The average continuum is shown in Fig. \ref{Fig:continuum}: the best fitting power-law index $\alpha_{\rm{steep}}=2.8\pm0.1$ is
consistent with the flare function decay $\propto t^{-2.7\pm0.1}$. Suggestively, this happens for $t\lesssim1000$ s: at later times the 
continuum is likely to be dominated by the shallow decay component instead of the steep decay emission (Figure \ref{Fig:continuum},
lower panel, shows a decreasing flare-to-continuum ratio around $t\sim 400$ s due to the progressively increasing contribution of the shallow
decay component to the continuum). During the first  $\sim1000$ s the \emph{typical flare-to-continuum ratio} is $L_{\rm{flare}}/L_{\rm{steep}}=4.7$
(median value) albeit with a large scatter. Using the data from C10 we obtain a similar median value $F_{\rm{flare}}/F_{\rm{steep}}=4.3$
(Fig. \ref{Fig:continuum}, upper panel, inset). The physical mechanism powering each flare emission is therefore required to release 
an average amount of energy $E_{\rm{flare}}$: 
\begin{equation}
	E_{\rm{flare}}(t)\sim  L_{\rm{flare}}	\Delta t\approx t L_{\rm{steep}}(t)
\end{equation}
where $\Delta t$ is the flare duration.  The median value $\Delta t/t=0.23$ from C10 has been used.

The \emph{average} continuum vs. flare function temporal behaviour of Fig. \ref{Fig:continuum} results from the presence of a correlation linking
the steep decay flux evolution to the flaring activity within \emph{individual} GRBs. Modelling the steep decay and the flaring activity of each burst
with decaying power-laws of index $\alpha_{\rm{steep}}$ and  $\alpha_{\rm{flare}}$, respectively, we obtain the result drawn in  
Fig. \ref{Fig:slopetoslope}\footnote{Only GRBs with multiple flares can  be used. Since the knowledge of the redshift is not necessary, 
we supplemented the present sample with additional 18 \emph{Swift}-GRBs with multiple flares and good data coverage
(red dots in Fig. \ref{Fig:slopetoslope}). These bursts are listed in Table \ref{Table:GRBsupplement}.}: flares seems to be linked 
to the contemporaneous steep decay flux evolution in a way that causes flatter flare luminosity functions to be associated to more gradual 
steep decays.

Generically speaking, these observations are consistent with a model where a first physical mechanism is responsible for the
steep decay continuum (without flares), while mechanism 2 powers X-ray flares. The presence of the $\alpha_{\rm{steep}}$ vs. 
$\alpha_{\rm{flare}}$ relation suggests that the two mechanisms are in some way related and that the sporadic appearance of 
mechanism 2 is triggered by some properties of mechanism 1. Instabilities affecting mechanism 1 can in principle provide the 
source of episodic releases of energy manifesting as flares. If this is the case, instabilities are likely to be  triggered by some 
physical quantity related to the decay of continuum flux at time $t$: this would explain why late-time flares are so rare 
(only $\sim5$\% GRBs show clear flaring activity for $t > 1000$ s), but also the paucity of flares in GRBs 
with simple power-law decay X-ray afterglows. In those GRBs, the steep decay is likely to be hidden by a contemporaneous 
but physically different emission component (see \citealt{Margutti10a} for a detailed analysis of the two emission components
in GRB\,081028).
We speculate that, given the remarkable similarity between X-ray flares and prompt pulses, it is possible that both mechanisms
also operate during the $\gamma$-ray prompt emission producing slowly varying (mechanism 1) and fast varying (mechanism 2)
emission components. The presence of both long (several seconds) and short ($\approx$ ms) variability time-scales in the same
burst is a known feature of the GRB prompt emission (e.g. \citealt{Norris96}, but also \citealt{Vetere06}, \citealt{Borgonovo07}).

L08 concluded that the outflow interaction with the stellar envelope cannot 
produce flares out of a continuous flow. In the following we analyse various mechanisms able to produce variability in the central 
engine release of energy 
in the context of accretion (Sect. \ref{SubSubSec:varaccretion}) and magnetar models (Sect. \ref{SubSubSec:magaccretion}).
\subsubsection{Variability in accretion models}
\label{SubSubSec:varaccretion}
In the context of GRB accretion models, a flare corresponds to a sudden increase in the jet luminosity as a consequence of
an abrupt change of the mass accretion rate\footnote{The efficiency $\eta_{\rm{acc}}\equiv\eta_{\rm{acc}}(a)$ of the process
is unlikely to undergo abrupt temporal changes (see e.g. Kumar et al. 2008a,b).}. Generally speaking, large $\dot m$ variations arise 
if the disk develops a ring-like structure: flares would be associated to the accretion of blobs of material initially located at various radii
which subsequently evolve on the viscous time scale. This class of models would naturally account for the observed duration-time scale
correlation  and duration- peak luminosity anticorrelation  (C10) as shown by \cite{Perna06}.
Different classes of disk instabilities could lead to this scenario\footnote{The importance of hydrodynamical instabilities in 
collapsar disks has been recently demonstrated by the studies of \cite{Taylor10}.}. 

Viscous
instability arises for $d\dot m/ d \Sigma <0$ where $\Sigma$ is the disk surface density (see e.g. \citealt{Franck85}). When this condition is
satisfied more material will be fed in those regions of the disk that are more dense while material will be removed from less dense areas, 
so that the disc will likely break up into rings. Viscous and thermal instabilities (the latter taking place for $dQ^{-}/dT<dQ^{+}/dT$
being $Q^{+}$ and $Q^{-}$ the heating and cooling rates respectively) have been invoked to explain Dwarf Novae outbursts:
in this case the instabilities give rise to a limit-cycle behaviour (see e.g. \citealt{Cannizzo98}) and do not lead to a total disk break down. 
While the $\dot m$ required to explain the Dwarf Novae outbursts luminosity 
is completely different from the conditions expected to hold in GRB hyper-accreting disks, we note that outside-in instability bursts (see 
\citealt{Franck85} and references therein) qualitatively share some observational properties with GRB X-ray flares/prompt pulses:
they have rapid rise -slower decay profiles and rise first at longer wavelength (see M10).  The stability conditions for
GRB hyper-accreting disks have been thoroughly studied: the \emph{steady state} disk model of \cite{DiMatteo02} led the authors to conclude that
the accretion flow is both thermally and viscously stable for a variety of $\dot m$ values supposed to give rise to the GRB phenomenology.
However, X-ray flares suggest that the GRB engine is long -lived: a \emph{time-dependent} computation is therefore required.
The time-dependent studies of \cite{Janiuk04} and \cite{Janiuk07} revealed the thermal-viscous instability to be an intrinsic 
property of the innermost disk radii for torus densities $\sim 10^{12}\,\rm{g\,cm^{-3}}$;
the GRB disk is not stabilised but rather breaks down into rings leading to several episodes of dramatic accretion on the viscous 
time-scale of each ring. However, large hyper-accretion rates are required for the instability to set in: $\dot m \gtrsim 10\, \rm{M_{\sun}\,s^{-1}}$.
This value is $\sim 2$ order of magnitude higher than the typical accretion rate invoked to explain the prompt GRB phase 
which is $\dot m_{\rm{prompt}}\sim 0.1-0.2\, \rm{M_{\sun}\,s^{-1}}$ (see \citealt{Lindner10} and references therein).

Both the black hole spin and the presence of large-structure magnetic fields can deeply modify the accreting torus structure: for a rapidly
spinning black hole, the disk-black hole magnetic coupling transfer rotational energy from the black-hole to the inner disk where
viscous-thermal instability is known to arise for a Schwarzschild black hole. The magnetic coupling results in reducing the $\dot m$
value for which thermal and viscous instabilities establish:  \cite{Lei09} showed that the disk becomes unstable in its inner region for 
accretion rates as low as $\dot m > 0.1\,\rm{M_{\sun}\,s^{-1}}$ ; \cite{Janiuk10} found that for a Kerr black hole instability in the inner 
edge may be  effective for $\dot m \sim 0.5\,\rm{M_{\sun}\,s^{-1}}$, the actual value depending on the viscosity parameter $\alpha$.
While a time-dependent analysis is required also in this case, these works prove that viscous and thermal instabilities could indeed
arise for $\dot m \approx \dot m_{\rm{prompt}}$.

Alternatively, the ring-like structure of the accretion disk could be the outcome of a gravitational instability occurring in the 
disk outskirts. Gravitational instability arises when $Q=\Omega c_{s}/\pi G \Sigma < 1$ (\citealt{Toomre64}) and is known
to lead to fragmentation in the outer regions of active galactic nuclei (see e.g. \citealt{Shlosman90})  and of young stellar 
objects (see e.g. \citealt{Adams93}). Hyper-accreting GRB disks in the steady state regime have been proven to undergo gravitational instabilities
when $\dot m \gtrsim 2\,\rm{M_{\sun}\,s^{-1}}$ and radii $r\gtrsim 30\,r_g$ ($r_g\equiv 2GM/c^2$) under the assumption of 
a viscous parameter $\alpha=0.1$. The threshold to gravitational instability lowers to $\dot m \gtrsim 0.2\,\rm{M_{\sun}\,s^{-1}}$, $r\gtrsim 15\,r_g$ 
for $\alpha=0.01$ (\citealt{Chen07}, but see also \citealt{DiMatteo02}). This mechanism has been suggested as possible source
of the GRB X-ray flares by \cite{Perna06} (see however \citealt{Proga06}; \citealt{Piro07} for a critical view).

Whatever the origin of the disk fragmentation\footnote{Note that in the \citealt{King05} picture is the stellar \emph{core} which 
undergoes fragmentation giving rise to a two-step collapse.} and/or ring-like structure is (i.e. viscous, thermal or gravitational), 
the total energy released by each flare is likely to be proportional to the total mass of the fragment $m_{\rm{f}}$ which gives rise to the burst 
of accretion. Writing $E_{\rm{flare}}\sim L_{\rm{flare}}\Delta t$, where $\Delta t$ is a measure of the flare duration, since
$L_{\rm{flare}}\propto t^{-2.7}$ and $\Delta t\propto t$ (C10), the average energy of a 
flare is found to scale as $E_{\rm{flare}}\propto t^{-1.7}$ for $30\,\rm{s}<t<1000\,\rm{s}$. In particular, using the best-fitting 
average flare luminosity calculated in Sec. \ref{Sec:simulations} and the flare width vs. peak time relation ($w\sim0.2\times t_{\rm{pk}}$) 
from C10 we find:
\begin{equation}
	\label{Eq:Eisoflare}
	E_{\rm{iso,flare}}(t)\approx1.3\times10^{54}\,(t/\rm{s})^{-1.7}\,(\rm{erg\,s^{-1}})
\end{equation}
where $E_{\rm{iso,flare}}$ refers to the isotropic energy emitted in the $1-10000$ keV energy band.
Equation \ref{Eq:Eisoflare} is valid in the time interval $40\,\rm{s}<t<1000\,\rm{s}$.
A multiplicative factor of 2 has been applied to account for the limited 2.2-14.4 keV X-ray energy band 
in which $L_{\rm{flare}}$ has been computed\footnote{The exact value of this factor depends on the values of the photon 
indices of the Band function describing the spectral energy distribution of flares. For the typical low energy and
high energy spectral photon indices $\alpha_{\rm{B}}=-1$, $\beta_{\rm{B}}=-2.5$ (\citealt{Kaneko06}), and spectral peak 
energy$E_{\rm{peak}}\sim$ a few keV, this factor is $\approx2$.}.
Our observations would therefore imply the fragments mass $m_{f}(t)\propto E_{\rm{flare}}(t)$ to scale as:
\begin{equation}
	\label{Eq:fragmentmass}
	m_f(t)\propto t^{-1.7}
\end{equation}
Since the arrival time $t$ of a blob of material initially at a distance $R$ is likely to scale as a positive 
power of the radius $t\propto R^{\beta}$ with $\beta>0$ ($\beta=3/2$ in the case of advection dominated flows, 
\citealt{Perna06}, their Eq. 2), Eq. \ref{Eq:fragmentmass} suggests that fragmentation taking place at larger radii and/or later times 
gives rise to lower mass objects. This is qualitatively consistent with 
\cite{Perna06}, their Eq. 5: however, the actual mass distribution of the fragments strongly depends on the local disk properties and can be
addressed only via numerical simulations (R. Perna, private communication). 
Multiple flares GRBs would correspond to a flatter $m_f$ vs. $t$ (and $m_f$ vs. $R$) dependence as implied by the results reported 
at the beginning of Sect. \ref{SubSec:var}. 

Yet another possibility giving rise to variability in accretion models is a modulation of the mass fall back
rate $\dot m_{\rm{fb}}$ as proposed by Kumar et al. 2008a,b: 
while the underlying physical mechanism has still to be understood, this mechanism is potentially able to give rise 
to sharp variations of $\dot m$ provided that $t_{\rm{acc}}\ll t$ (as shown by \citealt{Kumar08b}, their Fig. 6 and 7) and 
naturally accounts for  the $\propto t^{-2.7}$ behaviour as discussed in Sect. \ref{SubSubSec:slopeaccretion}. A possible 
realisation of this process is given by gravitational fragmentation of the fall back material into several bound objects
(\citealt{Rosswog07}).

Magnetic fields can suppress disk fragmentation  (\citealt{Banerjee06}), give rise to magneto-rotational instabilities (MRI)
and strongly modify the dynamics of accretion both in magnetars and black-hole systems (see e.g. \citealt{Zhang09}; \citealt{Proga03}
and references therein). In particular, a variable output may result from a modulation of the accretion rate by the magnetic-barrier 
and gravity (\citealt{Narayan03}). As mass is being accreted on to the black-hole, the magnetic flux is accumulated in the inner region, 
causing the accretion through the torus to be repeatedly stopped and then restarted. When accretion resumes, an X-ray flare
is produced (\citealt{Proga06}). According to \cite{Spruit05} the accumulated magnetic field $B$ can support the gas against 
gravity until the radial magnetic force $F_m$ is of the order of a few percent the gravitational force $F_g$. The force balance 
yields a minimum $B\propto \dot m^{1/2}$. Since the magnetic flux captured by accretion also depends on $\dot m$, then it 
is reasonable to expect the systems characterised by a flatter $\dot m(t)$ decay (manifested as a flatter afterglow steep decay) to be able 
to meet the $F_g\sim F_m$ requirement more easily and repeatedly (i.e. a number of flares are expected). 
This scenario would therefore naturally account for the $\alpha_{\rm{flare}}$ vs. $\alpha_{\rm{steep}}$ relation of Fig. \ref{Fig:slopetoslope}.
\subsubsection{Variability in magnetic models}
\label{SubSubSec:magaccretion}
This section concentrate on scenarios where the variability completely depends on magnetic effects. \cite{Giannios06} suggested that
flares can be powered by magnetic reconnection triggered by the deceleration with the external medium. However, the $\alpha_{\rm{flare}}$ vs. 
$\alpha_{\rm{steep}}$ relation (Fig. \ref{Fig:slopetoslope}), would require the steep decay to be intimately connected to the deceleration
phase of the fireball: \emph{Swift} observations instead favour an "internal" origin of the steep decay, being the steep decay either
interpreted as the tail of the prompt emission (e.g. \citealt{Willingale10} for a recent study) or the result of the prolonged engine activity
(e.g. Kumar 2008a,b). 

Alternatively, in the context of magnetar models \emph{differentially rotating} millisecond pulsars can provide an engine able to 
repeatedly store and release energy. Differential rotation causes toroidal magnetic fields to be repeatedly wound up to 
$\sim 10^{17}\,\rm{G}$ and then pushed to and through the pulsar surface by buoyant forces: this allows the neutron star spin energy 
to be emitted in powerful bursts of pulsar wind  (\citealt{Kluzniak98}; \citealt{Ruderman00}), an extreme but transient realisation 
of the  \cite{Usov92} pulsar.
This mechanism has been invoked to explain the flare phenomenology by \cite{Dai06}. The re-windup time, i.e., the time between 
subbursts is anticorrelated with the pulsar differential rotation $\Delta \Omega$; however, the luminosity of the subbursts (manifesting as
flares) is expected to roughly scale as $\approx \Omega(t) ^4$, being  $\Omega$ the neutron star rotation rate (\citealt{Ruderman00}, their Eq.
19 and 20). If the pulsar rotation $\Omega$ is also the ultimate source of energy which powers the steep decay, then it is 
not unreasonable to expect the correlation displayed in Fig. \ref{Fig:continuum} and Fig. \ref{Fig:slopetoslope}.
\section{Summary and conclusions}
\label{Sec:summary}
\begin{figure}
\vskip -0.0 true cm
\centering
    \includegraphics[scale=0.6]{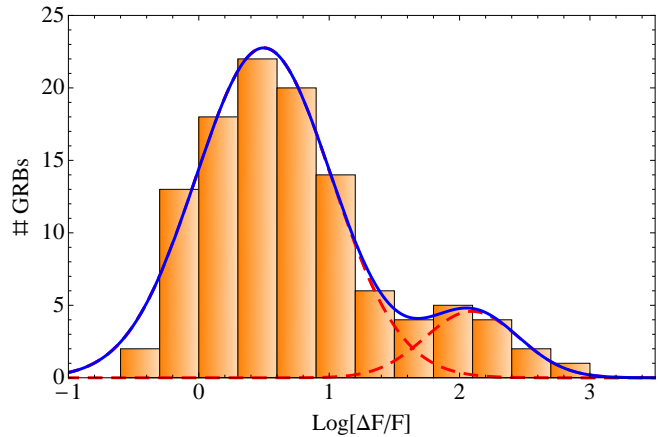}
      \caption{Flare peak flux to continuum ratio for the sample of early-time flares of C10. The distribution can be fitted by the
      superposition of two gaussian profiles with the following best fitting parameters: $\overline x_1=0.50 \pm 0.03$, $\sigma_1=0.52\pm0.03$;
      $\overline x_2=2.10 \pm 0.08$, $\sigma_2=0.36\pm0.09$.}
\label{Fig:fluxratio}
\end{figure}

We analysed the X-ray afterglows of 44 long GRBs observed by \emph{Swift} with the goal to determine the 
average flaring component in the common rest frame
energy band which is $2.2-14.4$ keV. No assumption on the flare functional shape has been made.   
Our work highlights the importance of the proper consideration of the threshold of detection of flares against the contemporaneous
continuous X-ray emission.  In particular we showed that:
\begin{itemize}
	\item	The best fit 2.2-14.4 keV average flare luminosity curve is: 
	$\langle L \rangle =10^{(54.5\pm0.1)} (t/\rm{s})^{-2.7\pm0.1}\,(\rm{erg\,s^{-1}})$ for $30\,\rm{s}<t<1000\,\rm{s}$ 
	(Fig. \ref{Fig:avelum}). 
	A similar scaling was obtained by C10 for the subsample of 43 early-time flares with redshift: 
	$L_{\rm{pk}}\propto t_{\rm{pk}}^{-2.7\pm0.5}$. The application  of the
	L08 method to this subsample of burst leads to $\langle L \rangle\propto t^{-2.6\pm0.1}$ (Fig. \ref{Fig:lazzatiplot}). 
	\item In the time interval $30\,\rm{s}<t<1000\,\rm{s}$ the typical 1-10000 keV flare isotropic energy is found to scale as:
	$E_{\rm{iso,flare}}\approx 1.3\times10^{54}\,(t/\rm{s})^{-1.7}\,(\rm{erg})$.
	\item For $t>1000$ s threshold effects related to the presence of a continuum X-ray emission underlying the erratic appearance
	of flares start to play an important role. The resulting $\langle L \rangle\propto t^{-1.2\pm0.1}$ is biassed towards the bright
	end of the flare luminosity distribution at these times. The un-biassed $\langle L \rangle$ is likely to be steeper (Fig. \ref{Fig:avelum}).
	\item According to the flare avoidance region analysis which properly accounts for the fraction of undetectable flares at any time $t$,
	the power-law decay index of $\langle L \rangle$ is steeper than $-1.8$ at the 90\% c.l. for $t<1000$ s (Fig. \ref{Fig:lumtimeplane}).
	\item GRBs with multiple-flare have a flatter than average flare luminosity function:  
	$\langle L \rangle\propto t^{-\alpha}$ with $0.6\lesssim\alpha\lesssim3$. Parenthetically, this is probably the reason why L08 determined
	a flatter average flare luminosity function $\langle L \rangle_{\rm{L08}}\propto t^{-1.5}$.
	\item The decay of the continuum closely tracks the decay of the average flare luminosity function (Fig. \ref{Fig:continuum}) for $t<1000$ s,
	suggesting that the two components are deeply related to one another. In particular, within individual GRBs, the power-law decay index of the 
	\emph{steep decay} is positively correlated to the power-law decay index of the flaring component (Fig. \ref{Fig:slopetoslope}). GRB\,100212A
	is a show case in this respect (\citealt{Grupe10}).
	The typical flare to steep-decay luminosity ratio is $L_{\rm{flare}}/L_{\rm{steep}}=4.7$.
	\item As a result, the typical flare energy at time $t<1000$ s obeys the following relation: 
	$E_{\rm{flare}}\sim L_{\rm{flare}}(t)\Delta t\approx t L_{\rm{steep}}(t)$ 
\end{itemize}
These findings suggest a model where the steep decay is produced by some form of activity of the internal engine which would be required to be 
still alive at  those times (see however \citealt{Genet09} for a complementary view), while flares could be powered by instabilities affecting 
the physical source of energy which gives origin to the steep decay. This would explain the X-ray flares erratic behaviour.
In this picture, the shallow decay phase would be due to  a completely distinct component of emission which progressively hides both the 
steep decay and the X-ray flares as time proceeds. 

The $\langle L \rangle\propto t^{-2.7\pm0.1}$ has been analysed in the context of accretion and magnetar models of GRBs. In particular:
\begin{itemize}
	\item	 According to the hyper-accreting black hole scenario, the $L\propto t^{-2.7}$ scaling can be obtained in the case of rapid accretion 
	($t_{\rm{acc}}\ll t$)  or when the last $\sim 0.5 M_{\sun}$ of the original $14 M_{\sun}$ progenitor star are accreted (Kumar et al., 2008a,b).
	Alternatively, the steep $\dot m\propto t^{-2.7}$ behaviour could be triggered by a rapid outward expansion of an accretion shock 
	in the material feeding a convective disk (\citealt{Lindner10}).
	\item The rotational energy of a rapidly spinning magnetar can be extracted to power a jet whose luminosity is likely to be between 
	the monopole ($L\propto e^{-2t}$) and dipole ($L\propto t^{-2}$) cases.  $L\propto t^{\sim-2.3}$ is expected, based on the braking indices 
	of observed pulsars in the Galaxy.
\end{itemize}

In both scenarios the variability, which is the main signature of the flaring activity, establishes as a consequence of different
kinds of instabilities. In the case of accretion models, thermal, viscous or gravitational instabilities could either lead to disk breakdown
or fragmentation. Our analysis constrains the mass of the accreting material to scale as $m_{f}(t)\propto t^{-1.7}$ (Eq. 7).  
However, the presence of magnetic fields gives rise to MRI instabilities and strongly modifies the dynamics of accretion:
the accumulation of magnetic flux during the accretion can repeatedly stop and restart the accretion process (\citealt{Proga06}). 
This would  account for the erratic flare emission while explaining the $\alpha_{\rm{flare}}$ vs. $\alpha_{\rm{steep}}$ relation.
Alternatively, differentially rotating millisecond pulsars provide a viable mechanism where the existence of the
$\alpha_{\rm{flare}}$ vs. $\alpha_{\rm{steep}}$ relation can be reasonably explained. We note that if the flare origin is linked
to the magnetic energy dissipation, the flare emission is likely to be polarised (\citealt{Fan05}), while a disk fragmentation origin
is likely to be accompanied by detectable gravitational wave signal (\citealt{Piro07}). Both signals will be detectable in the
near future.

Whatever the mechanism powering the X-ray flare emission is, it is extremely difficult to account for the late-time ($t>1000$ s) flare activity 
displayed by some bursts using the $\langle L \rangle \propto t^{-2.7}$ component: exceptional 
circumstances leading to the revival of the instabilities would be needed. An interesting possibility is offered by Fig. 
\ref{Fig:fluxratio}: while the lower edge of the distribution of the flux ratio is probably incomplete, this figure suggests the 
existence of two populations of X-ray flares (see also C10, their figure 13).
A first population with flux contrast $\sim5$ (which is the one responsible for the average flare and continuum behaviour of Fig. 
\ref{Fig:continuum}) and a second population of bright flares with a typical  $\Delta F/F\sim 100$ but extending up to 
$\Delta F/F\sim 1000$. Fig. \ref{Fig:continuum} directly links the X-ray flares to the underlying \emph{steep decay} flux.
It is therefore possible that at late times only the small fraction of flares belonging to the second population
are able to overshine the contemporaneous shallow decay component. This would explain why late time flares are so rare.
The detailed characterisation of the two flare populations is beyond the scope of the present work and is left for a future study.

\section*{Acknowledgments}
The authors thank the anonymous referee for constructive criticism. 
RM, RBD and RS thank Chris Lindner, Todd Thompson, Francesco Pasotti for valuable discussions.  
This work is supported by ASI grant SWIFT I/011/07/0, by the Ministry of University and Research of Italy 
(PRIN MIUR 2007TNYZXL), by MAE and by the University of Milano Bicocca, Italy.


\appendix
\section{Tables}
\setcounter{table}{0}

\begin{table}
\centering
\begin{minipage}{180mm}
 \caption{GRBs with multiple episodes of flaring activity (Fig. \ref{Fig:slopetoslope}).}
\begin{tabular}{lll}
\hline
GRB& GRB & GRB \\
\hline
050712  &060904A &081210\\
050713A &070107  &090831C\\
050726  &070129  &091026\\
050822  &070517  &100212A\\
060204B &070616  &100302A\\
060312  &080320  &100513A\\
\hline
\label{Table:GRBsupplement}
\end{tabular}
\end{minipage}
\end{table}

\label{lastpage}
\end{document}